\documentclass[aps,twocolumn,showpacs,preprintnumbers,superscriptaddress,groupedaddress,10pt]{revtex4-1}
\usepackage[utf8]{inputenc}
\usepackage{graphicx,amssymb,amsmath,amsthm,amsfonts,epsfig,times,natbib}

\usepackage[usenames,dvipsnames]{color}
\usepackage{epstopdf}
\usepackage{amsmath,amssymb}
\usepackage{tensor}
\usepackage{mathtools}
\usepackage{amsbsy}
\usepackage{bm,url}
\usepackage[linktocpage]{hyperref}

\usepackage[scaled]{beramono}
\usepackage[T1]{fontenc}

\definecolor{oxfordblue}{rgb}{0.0, 0.13, 0.28}
\definecolor{burgundy}{rgb}{0.5, 0.0, 0.13}
\definecolor{darkolivegreen}{rgb}{0.33, 0.42, 0.18}
\definecolor{darkblue}{rgb}{0,0,0.5}
\definecolor{richcarmine}{rgb}{0.84, 0.0, 0.25}
\definecolor{darkblue}{rgb}{0,0,0.5}
\definecolor{venetianred}{rgb}{0.78, 0.03, 0.08}
\definecolor{skobeloff}{rgb}{0.0, 0.48, 0.45}
\hypersetup{colorlinks=true, citecolor=darkblue, linkcolor=darkblue,
urlcolor = darkblue}

\newcommand{\ben}{\begin{enumerate}}
\newcommand{\een}{\end{enumerate}}

\def\be{\begin{equation}}
\def\ee{\end{equation}}
\def\bea{\begin{eqnarray}}
\def\eea{\end{eqnarray}}

\newcommand{\beq}{\begin{eqnarray}}
\newcommand{\eeq}{\end{eqnarray}} 
\newcommand{\ba}{\begin{align}}
\newcommand{\ea}{\end{align}}

\begin{document}

\title{Isolated black holes without $\mathbb{Z}_2$ isometry}

\author{
Pedro V. P. Cunha$^{1,2}$,
Carlos A. R. Herdeiro$^{1,2}$,
Eugen Radu$^{1}$
}

 \affiliation{${^1}$ Departamento de F\'isica da Universidade de Aveiro and CIDMA,
Campus de Santiago, 3810-183 Aveiro, Portugal}
\affiliation{${^2}$ CENTRA, Departamento de F\'isica, Instituto Superior
T\'ecnico, Universidade de Lisboa, Avenida Rovisco Pais 1,
1049 Lisboa, Portugal}
%


\date{October 2018}

\begin{abstract}
A mechanism to construct asymptotically flat, isolated, stationary black hole (BH) spacetimes with no $\mathbb{Z}_2$ (No$\mathbb{Z}$) isometry is described. In particular, the horizon geometry of such No$\mathbb{Z}$ BHs does not have the usual north-south (reflection) symmetry. We discuss two explicit families of models wherein No$\mathbb{Z}$ BHs arise. In one of these families, we exhibit the intrinsic horizon geometry of an illustrative example by isometrically embedding it in Euclidean 3-space, resulting in an ``egg-like" shaped horizon.  This asymmetry leaves an imprint in the No$\mathbb{Z}$ BH phenomenology, for instance in its lensing of light; but it needs not be manifest in the BH shadow, which in some cases can be analytically shown to retain a $\mathbb{Z}_2$ symmetry. Light absorption and scattering due to an isotropic source surrounding  a No$\mathbb{Z}$ BH endows it with a non-zero momentum, producing an \textit{asymmetry triggered BH rocket} effect.
\end{abstract}




\maketitle
\section{Introduction}
Knowledge of black hole (BH) physics is to some large extent based on exact solutions. Of central importance is the Kerr-Newman (KN) spacetime~\cite{Kerr:1963ud,Newman:1965my}, which according to the uniqueness theorems (see~\cite{Chrusciel:2012jk} for a review) is the most general, non singular (on and outside an event horizon) stationary, single BH solution of Einstein-Maxwell theory.  The spatial sections of the event horizon of KN BHs are, geometrically, squashed spheres~\cite{Smarr:1973zz}. These surfaces, moreover, have an antipodal isometry: the metric is invariant under a parity transformation $(\theta,\varphi)\rightarrow (\pi-\theta,\varphi+\pi)$ in the standard Boyer-Lindquist coordinates~\cite{Boyer:1966qh}. Due to the axi-symmetry of the solution this implies that the spacetime geometry, and in particular the horizon, has a $\mathbb{Z}_2$ symmetry: the northern and southern hemispheres are isometric. There is also an unambiguous equator corresponding to the set of fixed points of the $\mathbb{Z}_2$ isometry. An analogous (with adequate generalisations) antipodal isometry is widely acknowledged to be present in all known isolated BHs in all dimensions (see $e.g.$~\cite{Gibbons:2012ac}). 

These observations raise the following question: in four spacetime dimensions, can an isolated, asymptotically flat, stationary BH spacetime, free of singularities on and outside the event horizon, 
have a non $\mathbb{Z}_2$ isometric horizon? Here we show the answer is \textit{yes}, unveil a generic mechanism to construct non $\mathbb{Z}_2$ (No$\mathbb{Z}$) invariant BHs, and discuss how some
observables manifest the $\mathbb{Z}_2$ symmetry violation.

\section{Scalar deformations}
To search for non No$\mathbb{Z}$ BHs one must go beyond electrovacuum. The simplest additional matter content one may consider are  scalar fields. The existence of different examples of BHs with scalar hair, $e.g.$~\cite{Herdeiro:2014goa,Sotiriou:2014pfa,Herdeiro:2015waa,Doneva:2017bvd,Silva:2017uqg,Antoniou:2017acq,Herdeiro:2018wub}, some of which with rather different physical properties than KN (see $e.g.$~\cite{Cunha:2015yba}), justifies this choice. To establish a proof of principle, we consider deforming the KN geometry through a real scalar 
field $\phi$. 
Starting with the test field limit,  $\phi$ is  nonminimally coupled to the fixed KN background,
via some  tensorial scalar invariant source term  ${\cal J}$, constructed from the KN metric and gauge field. Its Lagrangian density is
\begin{eqnarray}
\label{lagsca}
\mathcal{L}_\phi=  
- \frac{1}{2}\partial_\mu \phi \partial^\mu\phi -f(\phi) {\cal J}(g;A) \ .
\end{eqnarray}
For concreteness we take $f(\phi)=e^{-2\alpha \phi}$, corresponding to an often considered non-minimal coupling, occurring naturally in, say, String Theory~\cite{Green:1987sp}, and Kaluza-Klein theory~\cite{Appelquist:1987nr}.  The specific choice of the constant $\alpha \neq 0$  is not central for our discussion. Below we shall also comment on other couplings. Employing the conventions in \cite{Townsend:1997ku}, the (dyonic) KN metric is $ds^2= \left[-{\Delta}(\omega_t)^2+{\sin^2\theta}(\omega_\varphi)^2\right]/{\Sigma}+{\Sigma}(dr^2/{\Delta} +d\theta^2)$, 
where $\omega_t\equiv dt-a\sin^2\theta d\varphi$,  $\omega_\varphi\equiv adt-(r^2+a^2)d\varphi$, $\Sigma\equiv r^2+a^2\cos^2\theta$, $\Delta\equiv r^2-2Mr+a^2+Q^2+P^2$ and $a\equiv {J}/{M}$. $(M,J;Q,P)$ are the ADM mass, angular momentum, electric and magnetic charges of the BH, respectively. The gauge connection is $
A= { Qr}
\left [\omega_t-P\cos\theta \omega_\varphi  
\right]/{\Sigma}$.

The scalar field equation derived from~\eqref{lagsca} is: 
\begin{eqnarray}
\label{eq}
\Box \phi+2\alpha e^{-2\alpha \phi} {\cal J}=0 \ .
\end{eqnarray}

We may now observe the following generic mechanism. For a $\mathbb{Z}_2$ invariant background, such as KN, the d'Alembertian $\Box$ operator is $\mathbb{Z}_2$ even.  But not all scalar invariants constructed from the KN metric and gauge field are $\mathbb{Z}_2$ even; some are $\mathbb{Z}_2$ odd and some are not $\mathbb{Z}_2$ eigenstates. 
Choosing one \textit{non}-$\mathbb{Z}_2$ even term as the source ${\cal J}$,
 the scalar field 
(and its energy-momentum tensor)
will be neither even nor odd under the $\mathbb{Z}_2$ transformation, and when considering its back reaction, if regular on and outside the horizon, it will lead to a No$\mathbb{Z}$ BH (with scalar hair).

%

\section{An electromagnetic source}
Let us illustrate this mechanism with an electromagnetic source. The simplest KN electromagnetic  invariant that is not a $\mathbb{Z}_2$ eigenstate is
\begin{eqnarray}
\label{source1}
 {\cal J}=\frac{1}{4} F_{\mu\nu} F^{\mu\nu}\equiv \frac{F^2}{4} \ ,
\end{eqnarray}
where $F_{\mu\nu}$ is the Maxwell tensor.  Indeed, for the KN solution:
\begin{equation}
\label{F2}
F^2=-\frac{16}{\Sigma^4}
\left[ 
b \left(\frac{\Sigma^2}{8}-\Sigma r^2+r^4\right) +  d \cos \theta  \left(\Sigma-2r^2\right)
\right] \ ,
\end{equation}
$b\equiv Q^2-P^2$, $d\equiv J P Q/ {M}$.
Whereas the first term in the square brackets is $\mathbb{Z}_2$ even, the second is $\mathbb{Z}_2$ odd. Thus, in the generic case with all $(M,J;Q,P)$ non-zero, the $F^2$ invariant is not an eigenstate of the $\mathbb{Z}_2$ isometry of the KN background $\theta \to \pi-\theta$. If this invariant sources a scalar field via~(\ref{eq}), such scalar field is not a $\mathbb{Z}_2$ eigenstate,  for any $\alpha \neq 0$. Moreover, this source allows non-singular solutions of the scalar field on and outside the event horizon, as confirmed by the existence of regular (fully non-linear)  Einstein-Maxwell-dilaton BHs~\cite{Garfinkle:1990qj}. 

\begin{figure}[h!]
\begin{center}
\includegraphics[width=0.27\textwidth]{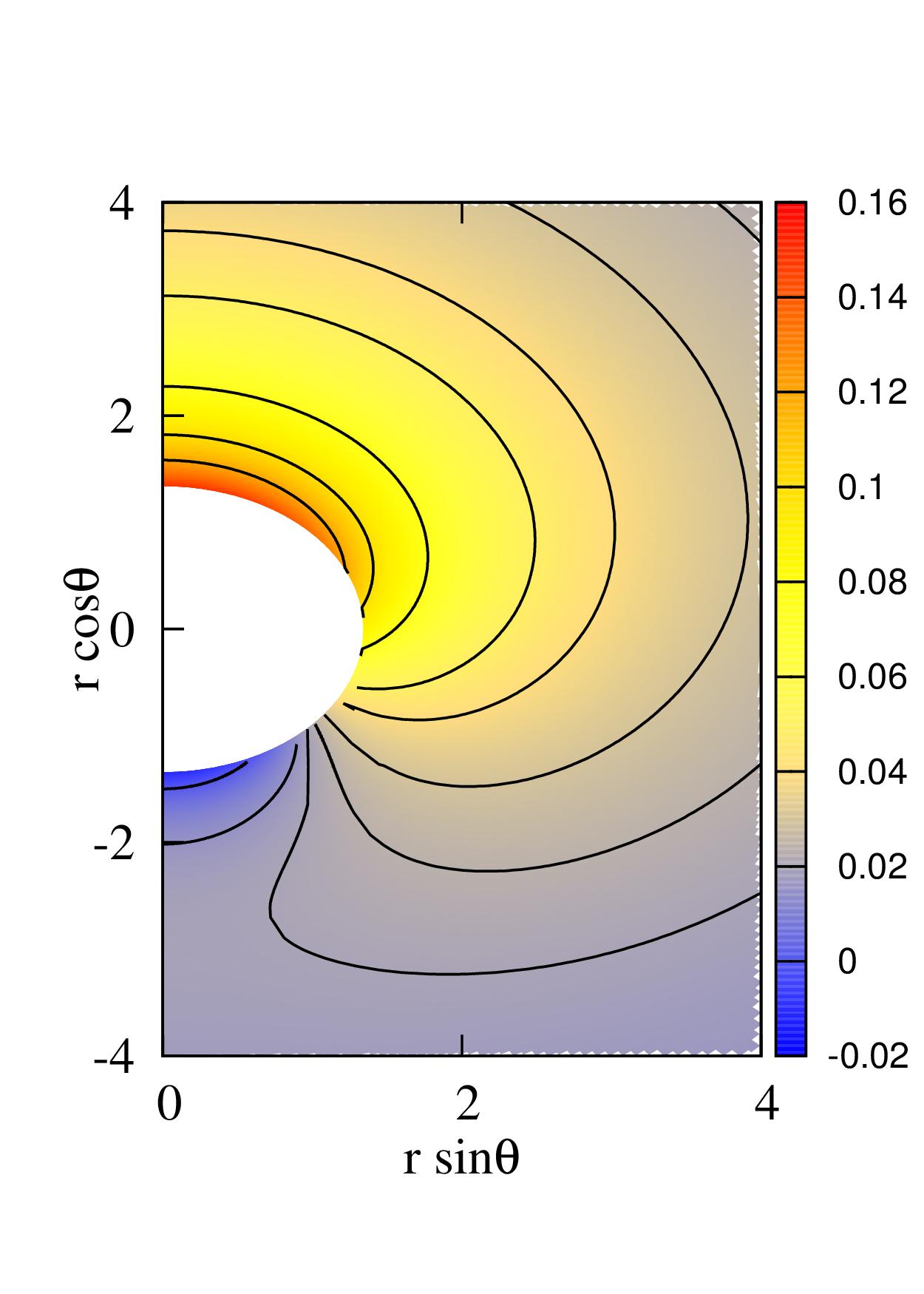}  \qquad 
\caption{\small{Scalar field amplitude obtained from~(\ref{eq}) with $\alpha=1$, using the  electromagnetic source~\eqref{source1} of a KN background. The contour lines are level sets and are clearly not $\mathbb{Z}_2$ invariant ($i.e.$ as $\theta \to \pi-\theta$).}}
\label{fig1}
\end{center}
\end{figure}

In Fig.~\ref{fig1} we plot the amplitude of the scalar field on a KN background with $(J;Q,P)=(0.6;0.4,0.6)$~\footnote{Here and below we use units with $M=1$.}. This was obtained by solving numerically~(\ref{eq}) with~\eqref{source1}-\eqref{F2} and $\alpha=1$. Such analysis confirms the solution is regular everywhere on and outside the horizon, vanishes asymptotically and it is not a  $\mathbb{Z}_2$ eigenstate, as it is clear from the plot.
 The failure to be a $\mathbb{Z}_2$ eigenstate can be analytically checked in the far-field, where the scalar field solution reads
$ \phi(r)={Q_s}/{r}+ (e+f \cos \theta)/{r^2}+\dots$;
the constants $Q_s, e,f$ depend on the background parameters. 
The sub-leading $1/r^2$ term, for instance, 
is generically 
not a  $\mathbb{Z}_2$ eigenstate.


The solution plotted in Fig.~\ref{fig1} is obtained in the test field limit. To establish the existence of No$\mathbb{Z}$ BHs the scalar field backreaction must be considered.  To obtain such fully non-linear solutions, and the corresponding deformed BHs, we couple the scalar field Lagrangian to the gravitational action. For simplicity, we use the electromagnetic source (\ref{source1}). Thus, we consider the model described  by the action
\begin{equation}
\label{actionKK}
\mathcal{S}= \int d^4 x \sqrt{-g} 
\left[
\frac{R}{4}-\frac{1}{2}\partial_\mu \phi \partial^\mu\phi  
-\frac{ e^{-2\alpha\phi}}{4} F_{\mu\nu} F^{\mu\nu}
\right] \ .
\end{equation}
For the main question here, qualitative difference are not expected for any $\alpha\neq 0$. Thus, as an example, we focus on the special case with $\alpha=\sqrt{3}$. This is the well known Kaluza-Klein Einstein-Maxwell-dilaton model, wherein the generalisation of the KN solution, a  rotating dyonic BH, is known~\cite{Rasheed:1995zv,Matos:1996km,Larsen:1999pp}
(see also \cite{Kleihaus:2003df}). 
This BH is characterised by the same four parameters $(M,J;Q,P)$ as the KN solution, but KN is not a special case of this BH (whereas the Kerr solution is).

According to our previous discussion, these BHs should fail to have (in the generic case, 
when all parameters are non-vanishing) a $\mathbb{Z}_2$ isometry. To establish this hitherto unnoticed fact,
it is enough to observe that the metric  and the matter functions
of this solution are combinations of building blocks of the generic form $U(r,\theta)=a_0+a_1 \cos \theta +a_2 \cos^2 \theta$, with
$a_i=a_i(r;M,J;Q,P)$.
As such, the solution is not, generically, $\mathbb{Z}_2$ isometric.

\begin{figure}[h!]
\begin{center}
\includegraphics[width=0.3\textwidth]{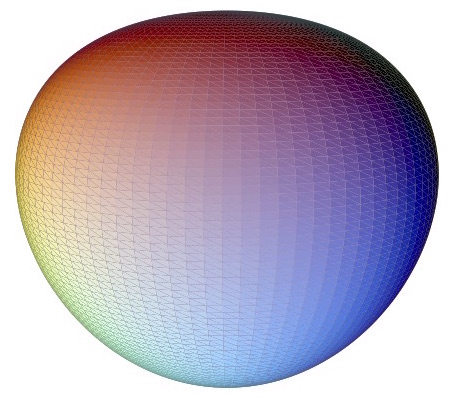} 
\includegraphics[width=0.4\textwidth]{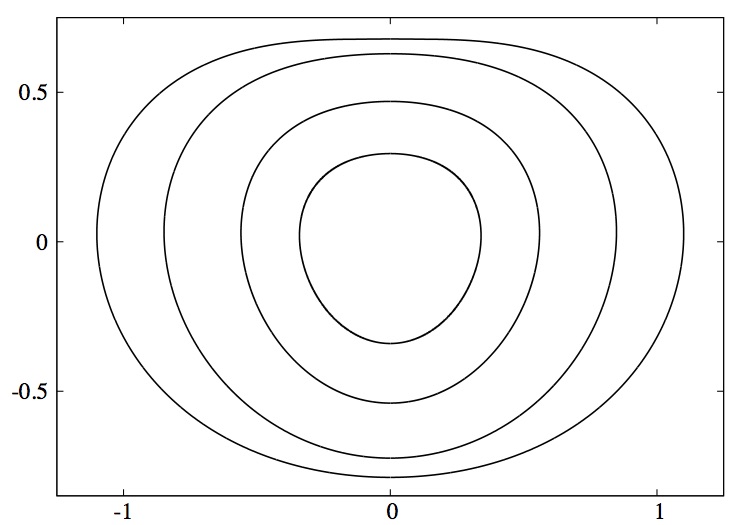} 
\caption{\small{Euclidean embeddings of the intrinsic horizon geometry for rotating dyonic BHs in Kaluza Klein theory.  (Top panel) A 3D plot for the solution with $(J;Q,P)=(0.035;1.87,0.01)$. (Bottom panel) 2D plots (constant azimuthal coordinate) for a sequence of solutions with $(J;Q,P)$ varying between $(0.54;0.23,1.01)$ (largest) and $(0.065;0.02,1.8)$ (smallest).}}
\label{fig2}
\end{center}
\end{figure}

We can now focus on the horizon geometry of one of these BHs 
\footnote{The explicit form of the near horizon extremal solution can be found in Ref. \cite{Astefanesei:2006dd}}. 
The spatial sections of the event horizon have an induced metric of the form 
$d\sigma^2=g_{\theta \theta}(\theta) d\theta^2+[g_{\theta \theta}(0)]^2
\sin^2 \theta d\varphi^2/g_{\theta \theta}(\theta)$,
where $ g_{\theta \theta}(\theta) =  \sqrt{ (\sum_{i=0}^2 b_i \cos^i\theta) (\sum_{i=0}^2 c_j \cos^j\theta})$, 
and  $b_i,c_i$ have cumbersome expressions in terms of $(M,J;Q,P)$. 
It is obvious that no  $\mathbb{Z}_2$ isometry exists in  the generic case, the usual north-south symmetry being lost.
This is illustrated in Fig.~\ref{fig2}, 
where  the embedding of the induced horizon metric in Euclidean 3-space of some solutions are shown. 
The construction of these embeddings follows a standard procedure~\cite{Smarr:1973zz}. 
One observes this intrinsic geometry has an ``egg-like" shape (see also~\cite{Kleihaus:2004} for a related discussion).


\section{A gravitational source}
As a second model wherein No$\mathbb{Z}$ BHs occur, we again start with a test field analysis, 
but   specialise  for simplicity 
the KN background to the uncharged limit (Kerr, $Q=0=P$) and consider a scalar-tensor model (no Maxwell field). Whereas the Kretschmann invariant, $R^{\mu\nu\alpha\beta}R_{\mu\nu\alpha\beta}$, is $\mathbb{Z}_2$ even, thus unsuitable to be the source for the purpose in sight, the Pontryagin density is $\mathbb{Z}_2$ odd. Consequently, we take the latter as the source term:
\begin{equation}
\label{source2}
 {\cal J}=
 \xi {\,^\ast\!}R^\mu{}_\nu{}^{\alpha \beta} R^\nu{}_{\mu \alpha \beta}~, \ \ \ \ 
~~{^\ast}R^\mu{}_\nu{}^{\alpha\beta}\equiv \frac12 \epsilon^{\alpha\beta \sigma \tau}R^\mu{}_{\nu \sigma \tau}~,
\end{equation}
where $\xi$ is a dimensionful coupling constant ($[\xi]=$Length$^2$), and $\epsilon^{\alpha\beta \sigma \tau}$ are the components of the Levi-Civita tensor.  
For Kerr, the  Pontryagin density is compactly written as
\begin{equation}
\label{action2}
{\,^\ast\!}R^\mu{}_\nu{}^{\alpha \beta} R^\nu{}_{\mu \alpha \beta}=\frac{96 a M^2  r \cos \theta }{\Sigma^6}
\left(
3\Sigma^2-16 r^2 \Sigma+16 r^4
\right)~,
\end{equation}
confirming it is $\mathbb{Z}_2$ odd.
Again, this source allows non-$\mathbb{Z}_2$ symmetric, non-singular solutions of the scalar (test) field on and outside the event horizon,  similar to those in Fig.~\ref{fig1}.

In this example, unlike the case of the electromagnetic source, fully non-linear solutions in closed analytic form are unknown and, likely, do not exist. Such solutions can, however, be constructed numerically, following~\cite{Delsate:2018ome}, in the corresponding non-linear model:
\begin{equation}
\label{actionCS}
\mathcal{S}= \int d^4 x \sqrt{-g} 
\left[
\frac{R}{4}
-\frac{\partial_\mu \phi \partial^\mu\phi}{2}
- \frac{\xi e^{-2\alpha\phi}}{4}   {\,^\ast\!}R^\mu{}_\nu{}^{\alpha \beta} R^\nu{}_{\mu \alpha \beta}
\right] \ .
\end{equation}
This model is a variation of the usual dynamical Chern-Simons modified gravity, wherein the coupling of the scalar field to the  Pontryagin density is of the form $\phi  {\,^\ast\!}R^\mu{}_\nu{}^{\alpha \beta} R^\nu{}_{\mu \alpha \beta}$~\cite{Jackiw:2003pm,Alexander:2009tp}. In the latter, therefore, the source term~\eqref{action2} gives rise to a $\mathbb{Z}_2$ odd scalar field, 
which is compatible with a $\mathbb{Z}_2$ even geometry
\cite{Yunes:2009hc,Yagi:2012ya,Stein:2014xba}. 

We have confirmed that the $\mathbb{Z}_2$ symmetry is lost for BH solutions of~\eqref{actionCS} 
with any $\alpha\neq 0$. Using the same coordinate system as in~\cite{Delsate:2018ome}, 
we define the ``equatorial" plane  as corresponding
to the value of $\theta=\theta_0$
which maximizes the proper length $L_\theta=\int_0^{2\pi} d\varphi \sqrt{g_{\varphi \varphi}(r_H,\theta)}$ of a $\theta=$const. circle on the induced horizon metric.
For the KN metric (or the solutions in \cite{Delsate:2018ome}),
$L_\theta$ is maximized for $\theta_0=\pi/2$.
This is not the case for the generic BH solutions of~\eqref{actionCS}.
We define the following measure
for the $\mathbb{Z}_2$ symmetry  violation $\epsilon \equiv 1-{L_p^{(N)}}/{L_p^{(S)}}$, 
where $L_p^{(N)}$ ($L_p^{(S)}$) is the proper length from the north (south)
pole to the ``equatorial" plane, $
L_p^{(N)}=\int_0^{\theta_0} d\theta \sqrt{g_{\theta \theta}(r_H,\theta)}$,
$L_p^{(S)}=\int_{\theta_0}^{ \pi} d\theta \sqrt{g_{\theta \theta}(r_H,\theta)}$.

\begin{figure}[h!]
\begin{center}
\includegraphics[width=0.4\textwidth]{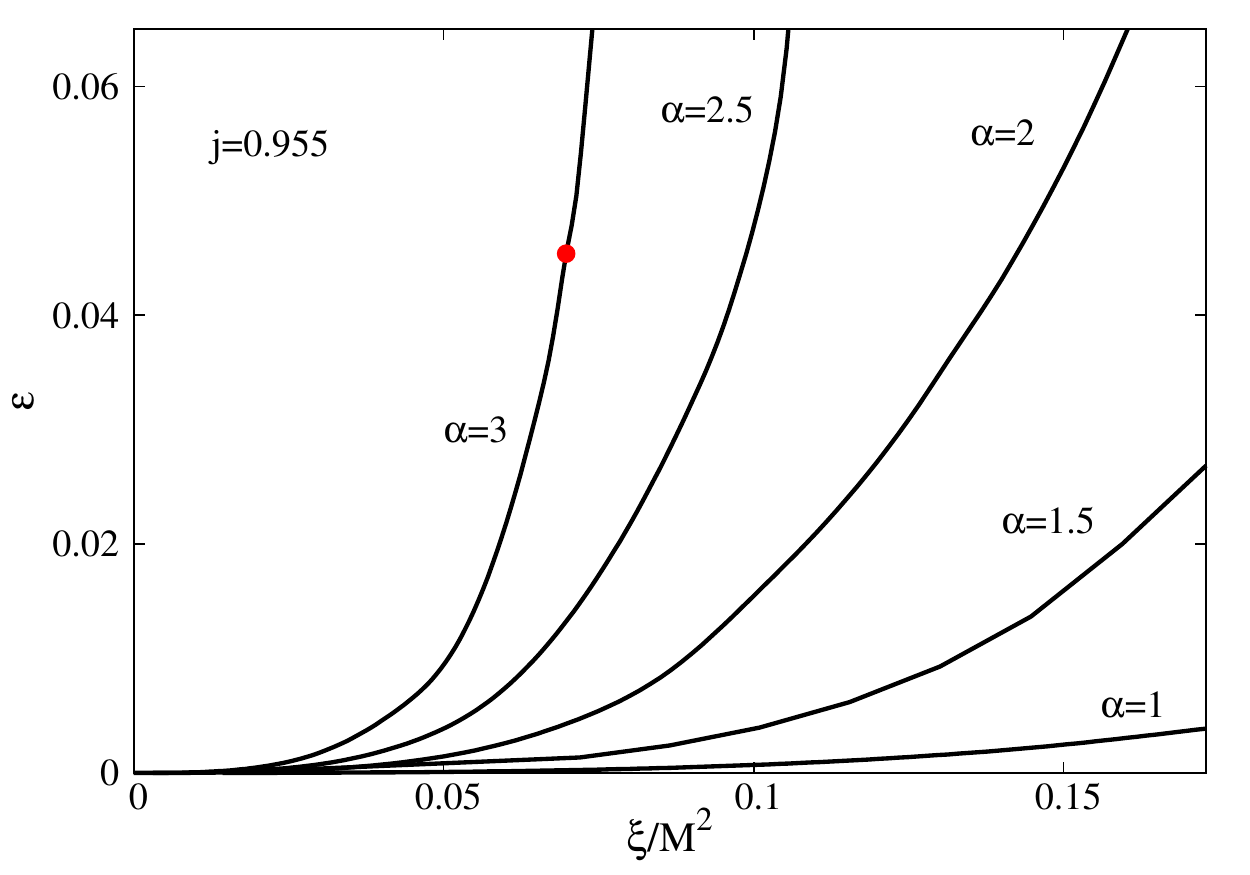} 
\includegraphics[width=0.27\textwidth]{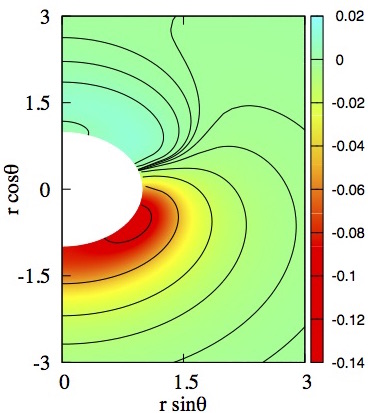} 
\caption{\small{(Top panel) Deformation $\epsilon$ $vs.$ the dimensionless parameter
$\xi/M^2$ for several values of $\alpha$. (Bottom panel) Scalar field level sets for the solution highlighted as a dot in the upper panel.}}
\label{fig3}
\end{center}
\end{figure}

In Fig.~\ref{fig3} (top panel) we exhibit the deformation $\epsilon$ for a subset of solutions of~\eqref{actionCS}.
In the region where these numerical solutions are enough accurate,
$\epsilon$ is no larger than a few percent, and the most significant  $\mathbb{Z}_2$ symmetry violations are found for near extremal BHs. By comparison, $\epsilon\sim 0.2$ for the solution in Fig.~\ref{fig2}. Thus, within the domain analysed, $\mathbb{Z}_2$ deformations within model~\eqref{actionCS} are barely visible in an embedding diagram. Still, the $\mathbb{Z}_2$ symmetry violation is clear, as shown in Fig.~\ref{fig3} (bottom panel) where the contour lines of the scalar field amplitude are shown for a particular, \textit{fully non linear} solution. We anticipate BHs larger values of $\epsilon$ exist in this model. Their construction, however, is challenging.

\section{Phenomenology}
\begin{figure}[h!]
\begin{center}
\includegraphics[width=0.35\textwidth]{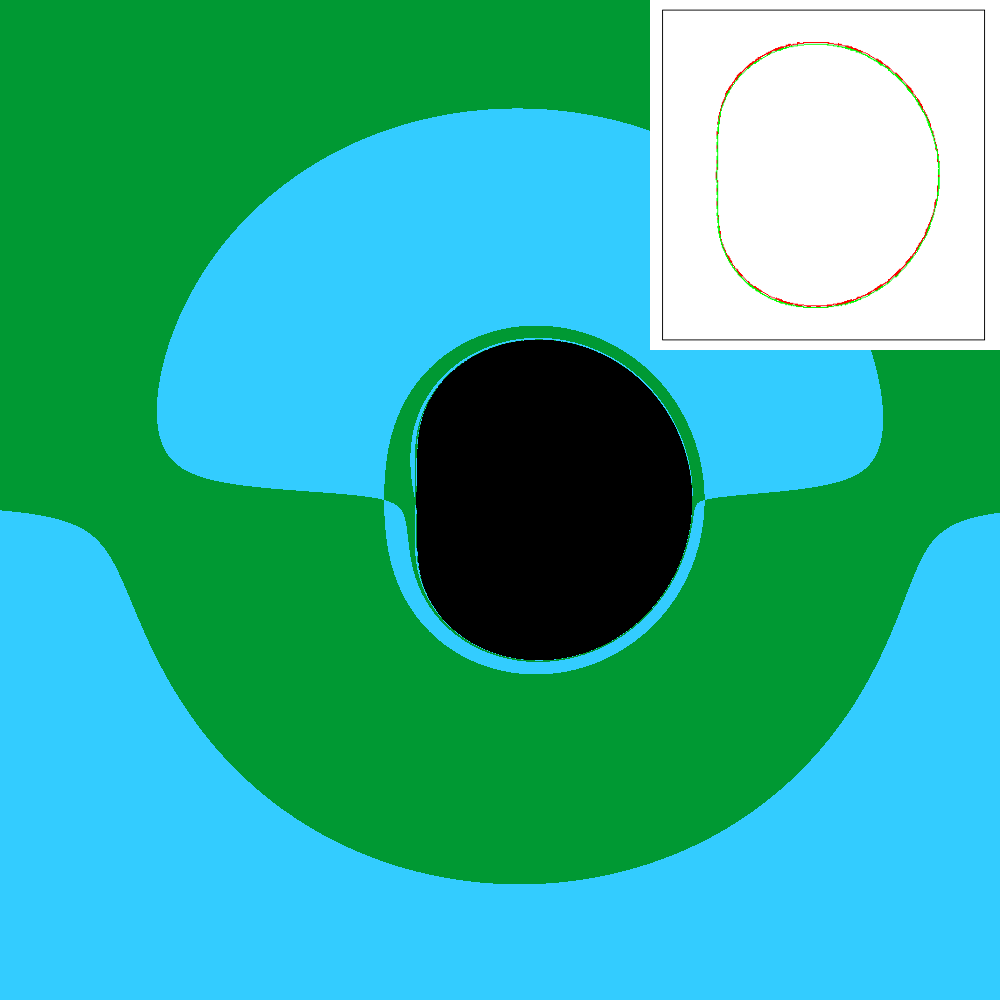} 
\caption{\small{Lensing and shadow due to a No$\mathbb{Z}$ dyonic BH with $(J,Q,P)\simeq (0.22,0.15,1.5)$. The green (light blue) colour light is emitted by the north (south) far away celestial sphere. The image is for an observer on the $\theta=\pi/2$ plane. The inset shows the shadow contour and its $\mathbb{Z}_2$ reflection. Both curves coincide. Thus, despite the lack of $\mathbb{Z}_2$ of the lensing, the shadow is $\mathbb{Z}_2$ symmetric.}}
\label{Fig-lensing}
\end{center}
\end{figure}

What could be the phenomenological impact of the absence of a $\mathbb{Z}_2$ isometry? A simple diagnosis can be performed using null geodesics as phenomenological probes. To maximise the $\mathbb{Z}_2$ violation effects we analyse model~\eqref{actionKK} rather than~\eqref{actionCS}. We have performed ray tracing (following~\cite{Cunha:2015yba}, see also~\cite{Amarilla:2013sj}) in an illustrative No$\mathbb{Z}$ dyonic rotating BH, and exhibit in Fig.~\ref{Fig-lensing} its gravitational lensing. 

Each point's colour in Fig.~\ref{Fig-lensing} encodes the geodesics's initial position: green (light blue) color denote geodesics that have their origin on the north (south) hemisphere of a far-away celestial sphere, enclosing both the BH and the observer. The first relevant feature of the image is that, even though the observer is placed on the (would be) equatorial plane surface $\theta=\pi/2$, the colored pattern is not $\mathbb{Z}_2$ symmetric, as can be apparent by interchanging the green/light blue colors. This further implies that $\theta=\pi/2$ is not a totally geodesic sub-manifold, as it would be in the case of spacetimes with a $\mathbb{Z}_2$ isometry. The black region in Fig.~\ref{Fig-lensing} is associated to null geodesics that would have their origin on the event horizon surface, and forms the BH \textit{shadow}~\cite{Falcke:1999pj}. The latter is a direct probe of the geometry close to the BH, wherein the fundamental photon orbits are found~\cite{Cunha:2017eoe}; it is an observable of ongoing astronomical observations~\cite{Loeb:2013lfa,Goddi:2017pfy}. 

The second (surprising) relevant feature in Fig.~\ref{Fig-lensing} is that the shadow edge displays a $\mathbb{Z}_2$ reflection symmetry, as illustrated by the inset of Fig.~\ref{Fig-lensing}. This can be shown analytically. The Hamilton-Jacobi equation for null geodesics turns out to be fully separable for these dyonic BHs. Liouville integrability for \textit{null} geodesics follows from the existence of a non-trivial fourth Carter-like constant of motion $K$, associated to a \textit{conformal} Killing tensor~\cite{Aliev:2013jya, Woodhouse1975}.  The null geodesic equations for the $\theta$-sector can be put in the form $h(p_\theta^2,\,K,\,{L}/{E},\,\theta)=0$, where $p_\mu$ is the 4-momentum, and $E\equiv-p_t$, $L\equiv p_\varphi$. Given an observer with fixed $\theta$, if a detected geodesic $\{p_\theta,K,{L}/{E}\}$ is part of the shadow, then the geodesic with $\{-p_\theta,K,{L}/{E}\}$ must also be part of the shadow, as a consequence of the $h$ functional dependence on $p_\theta^2$, and the radial sector independence on this transformation. Since the vertical axis of Fig.~\ref{Fig-lensing} is proportional to $p_\theta$, this implies that the shadow must be $\mathbb{Z}_2$ symmetric, regardless of the observation angle $\theta$. This shadow symmetry is non trivial; indeed the spherical photon orbits responsible for the shadow edge~\cite{Teo2003,Bardeen1973, Cunha:2017eoe,Cunha:2018gql} are in this case (generically) not $\mathbb{Z}_2$ symmetric with respect to $\theta=\pi/2$, in contrast to the Kerr case. This example sharply illustrates that the BH shadow is not a faithful probe of the event horizon geometry~\cite{Cunha:2018gql}.

Yet, the $\mathbb{Z}_2$ symmetry of the shadow (at any observation point) does not guarantee the overall  light absorption by the BH is north-south symmetric, since the  shadow seen at the north and south poles has a different size (an analogous effect should occur for time-like particles).  As a consequence of this the BH may acquire a {\it thrust} due to the asymmetric momentum absorption, 
leading to {\it a BH rocket}. 

\begin{figure}[h!]
\begin{center}
\includegraphics[width=0.38\textwidth]{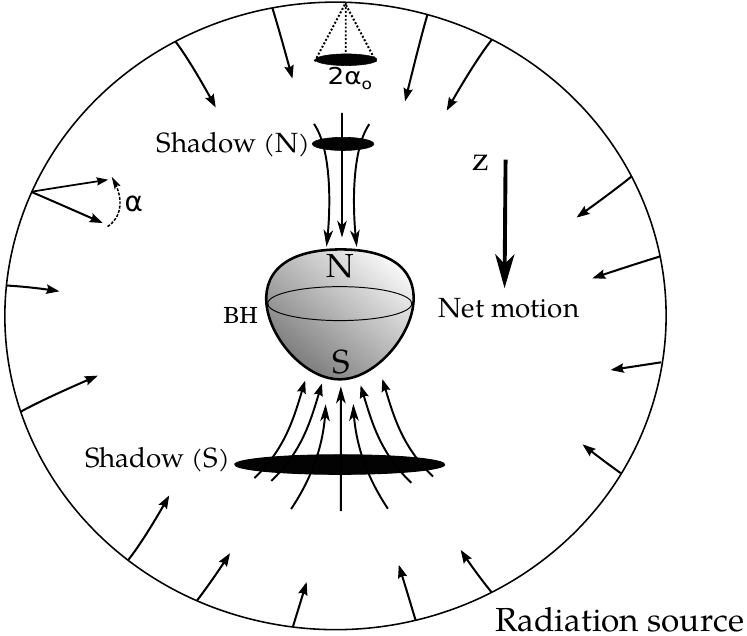} 
\caption{\small{Illustration of the thrust imparted on a No$\mathbb{Z}$ BH due to an isotropic light source (celestial sphere).}}
\label{infall}
\end{center}
\end{figure}

Consider that each point of the celestial sphere is a (quasi)-isotropic radiation source of light rays with $L=0$. Each light ray is parametrised by the inclination angle $\alpha\in[-\pi/2, -\pi/2]$, with $\alpha=0$ pointing to the BH center - Fig.~\ref{infall}. 
It can be shown, as detailed in the Appendix, that the $z$-momentum flux of radiation \textit{absorbed} by the BH is asymmetric. The flux contribution from \textit{scattered} radiation (not absorbed by the BH) is also asymmetric, although larger and with opposite sign. For the BH in Fig.~\ref{fig2} the total flux supplied by the south hemisphere is $\sim 3\%$ smaller than the north one, triggering a ``thrust" in the negative $z$ direction, $cf.$ Fig.~\ref{infall}.

\section{Final Remarks}
In the presence of a negative cosmological constant  static BHs exist without any spatial continuous symmetry~\cite{Herdeiro:2016plq}, since asymptotically Anti-de-Sitter spacetimes allow boundary conditions that anchor horizon deformations. The construction herein shows that non-minimal couplings allow asymptotically flat, isolated, regular on and outside an event horizon, stationary BHs with uncommon deformations, such as the loss of the usual north-south $\mathbb{Z}_2$ isometry. A similar construction can be made taking the scalar coupling   $f(\phi)$ in~\eqref{lagsca} to be a more general function. In particular, if this is a function of $\phi^2$ only, then KN (or Kerr) is also a solution of the fully non-linear model, coexisting with the scalarised BH~\cite{Doneva:2017bvd,Silva:2017uqg,Antoniou:2017acq,Herdeiro:2018wub},
which is not the case in the models~\eqref{actionKK}, \eqref{actionCS}.

\bigskip

\section*{Acknowledgements}
We would like to thank T. Tanaka and C. Yoo for a valuable discussion on lensing and in particular, suggesting the disk analogy. P.C.  is  supported  by
Grant No.  PD/BD/114071/2015 under the FCT-IDPASC Portugal Ph.D. program. C. H. and E.R. acknowledge funding from the FCT-IF programme and the FCT grant PTDC/FIS-OUT/28407/2017.  This work was supported by the European  Union's  Horizon  2020  research  and  innovation  programme  under  the H2020-MSCA-RISE-2015 Grant No.   StronGrHEP-690904, the H2020-MSCA-RISE-2017 Grant No. FunFiCO-777740  and  by  the  CIDMA  project UID/MAT/04106/2013. The authors  would also  like  to  acknowledge networking support by the COST Action GWverse CA16104. Computations were partly performed at the cluster ``Baltasar-Sete-S\'ois" and supported by the H2020 ERC Consolidator Grant ``Matter and strong field gravity: New frontiers in Einstein's theory" grant agreement no. MaGRaTh-646597."

\appendix


\section{Absorbed flux and black hole rocket effect}
\label{apA}
As in the main text, consider that each point of the celestial sphere is a (quasi)-isotropic radiation source of light rays with $L=0$. Each light ray is parametrised by the inclination angle 
\begin{equation}
\alpha\in[-\pi/2, -\pi/2] \ ,
\end{equation}
with $\alpha=0$ pointing to the BH center, $cf.$ Fig.~\ref{infall}.

\bigskip

The $z$-momentum flux $P_z^{\rm em}$ \textit{emitted} at the source is 
\begin{equation}
P_z^{\rm em}=\zeta\int {\frac{dP_z}{dN}}\,d\alpha\,dA \ ,
\end{equation}
where $\zeta={d^2N/(dA\,d\alpha)}$ is the (constant) flux density of photons per unit angle and area, and $dA=R^2\sin\theta\,d\theta d\varphi$ is the area element of the emitting sphere at large radius $R$. 
By symmetry 
\begin{equation}
P_z^{\rm em}=0 \ .
\end{equation}

\bigskip

To compute the $z$-momentum flux $P_z^{\rm {ab}}$ {of radiation \textit{absorbed}} by the BH take   
\begin{equation}
{\frac{dP_z}{dN}}=-\mathcal{P}\,\left(\cos\alpha\,\cos\theta-\sin\theta\,\sin\alpha\right)\,\Theta(\alpha) \ ,
\end{equation}
where $\mathcal{P}$ is the {initial} momentum per photon, and $\Theta$ is a step function which is 1 (0) if $\alpha$ is part (not part) of the BH shadow. Since the shadow is symmetric for every static observer the step function restricts the $\alpha$ domain into  $\alpha\in[-\alpha_o,\alpha_o]$, where $\alpha_o(\theta)$ computes the shadow size for each $\theta$. 

The BH absorbed momentum is then 
\begin{equation}
P_z^{\rm {ab}}=-(4\pi R^2\zeta\mathcal{P})\int_0^\pi d\theta\,\sin\theta\cos\theta\sin\alpha_o \ ,
\end{equation}
which, in general, may be non-vanishing. {Depending on the type of surface emission and how the $L=0$ light ray selection is implemented, the previous integral could have additional powers of $\cos\alpha_o$. However, for large $R$ one has $\cos\alpha_o\simeq 1$, leading to the same integral}. Using geodesic integrability and defining $u\equiv \cos\theta$  yields 
\begin{equation}
P_z^{\rm {ab}}=\int_{-1}^1 du \chi \ , \end{equation} 
where 
\begin{equation}
\chi\equiv -(4\pi R\zeta\mathcal{P})\,u\,\sqrt{K-q_1u-q_2u^2} \ ,
\end{equation}
$K$ is the Carter constant of the $L=0$ spherical photon orbit, and $q_1,q_2$ are geodesic separation constants of the $r$ and $\theta$ sectors. If $q_1=0$, then $P_z^{\rm {ab}}=0$ due to the $\mathbb{Z}_2$ reflection symmetry. 

Defining the north and south fluxes as 
\begin{equation}
P_N^{\rm {ab}}=\int_0^1\,du \chi \ , \qquad   P_S^{\rm {ab}}=\int_{-1}^0\,du \chi \ ,
\end{equation} for the BH in Fig.~\ref{fig2}, $|P_S^{\rm {ab}}|$ is $\sim 7\%$ larger than $|P_N^{\rm {ab}}|$. {However, these calculations only concerned radiation that was absorbed by the BH. \textit{Scattering} radiation that never falls into the BH can also provide a contribution $P_z^{\rm scat}\neq 0$ to the total $z$-momentum flux that is transferred to the BH. Numerically, one typically has $2\sim-(P_z^{\rm scat}/P_z^{\rm ab})$, $\,\,i.e.$ the scattering contribution is almost double (in modulus) but with the opposite sign, thus triggering a BH ``thrust" in the negative $z$ direction, $cf.$ Fig.~\ref{infall}. 
This result is consistent with the contribution of the north and south hemispheres ($P^T_N$ and $P^T_S$ respectively) to the total flux $P^T=P_z^{\rm scat} + P_z^{\rm ab}$. For instance, considering the BH in Fig.~\ref{fig2}, $P^T_S$ is $\sim 3\%$ smaller than $P^T_N$.}\\

{To obtain some insight into this result, consider the following analogy : take a two-sided disk with a totally reflecting surface on the top side (facing $z>0$) and a totally absorbent surface on the bottom side (facing $z<0$). A photon with $z$-motion which is reflected on the top side contributes $-2\mathcal{P}$ to the disk's $z$-momentum, whereas a photon absorbed in the bottom side contributes $+\mathcal{P}$; as a result the disk is endowed with a (net) negative $z$-momentum.}

\bibliography{letter_essay2018}

 
\end{document}